\documentclass[pdflatex,sn-mathphys-num]{sn-jnl}

\usepackage{graphicx}%
\usepackage{multirow}%
\usepackage{amsmath,amssymb,amsfonts}%
\usepackage{amsthm}%
\usepackage{mathrsfs}%
\usepackage[title]{appendix}%
\usepackage{xcolor}%
\usepackage{textcomp}%
\usepackage{manyfoot}%
\usepackage{booktabs}%
\usepackage{algorithm}%
\usepackage{algorithmicx}%
\usepackage{algpseudocode}%
\usepackage{listings}%
\usepackage{placeins}
\usepackage{subcaption}
\usepackage{multirow}
\usepackage{multicol}
\usepackage[normalem]{ulem}

\begin{document}

\title[Article Title]{Characterization of Rocking Block Behaviors with Classical and Alternative Restitution Models}

\author*{\fnm{Fernando} \sur{Gaibor E.}}\email{fdgaibor@espol.edu.ec}

\author{\fnm{Alexander} \sur{López}}\email{alexlop@espol.edu.ec}

\author{\fnm{Esther D.} \sur{Gutiérrez}}\email{egutierr@espol.edu.ec}

\affil{\orgdiv{Facultad de Ciencias Naturales y Matemáticas},
\orgname{Escuela Superior Politécnica del Litoral},
\orgaddress{\street{Campus Gustavo Galindo Velasco, Km 30.5 Vía Perimetral},
\city{Guayaquil},
\postcode{090112},
\state{Guayas},
\country{Ecuador}}}

\abstract{
This work investigates how restitution modeling affects the dynamics of rocking blocks subjected to harmonic excitation. While several studies have reported discrepancies between experimentally observed impact behavior and the predictions obtained using the classical Housner restitution coefficient, the implications of adopting alternative restitution formulations on the global dynamics of rocking systems remain largely unexplored. The system is formulated as a hybrid non-smooth dynamical model and analyzed through bifurcation diagrams, Lyapunov exponents, and basins of attraction for different slenderness ratios.

By comparing the classical restitution model proposed by Housner with the alternative formulation of Mao et al., we show that the choice of restitution model strongly influences the predicted system response. The alternative formulation leads to an earlier onset and greater prevalence of complex oscillations, as well as changes in the type, stability, and accessibility of attractors compared to the classical model. However, as the slenderness ratio increases, the dynamical features produced by both formulations progressively converge, indicating a reduced sensitivity to the restitution model for taller blocks. These results provide a dynamical perspective on why alternative restitution formulations, which predict impact responses closer to experimental observations, can produce markedly different behaviors from those obtained using the classical Housner model.
}


%%\pacs[JEL Classification]{D8, H51}

%%\pacs[MSC Classification]{35A01, 65L10, 65L12, 65L20, 65L70}

\maketitle

\section{Introduction}
Rocking blocks have been studied for decades due to their relevance in modeling and protecting structures such as masonry walls \cite{MasonryWalls2017}, statues \cite{AssymetricGeometry_4}, art pieces \cite{3DGeometry_2}, elevated water tanks \cite{WaterTank}, and other rigid-block-like objects requiring seismic protection \cite{RB_SP_NLD2}.
The fundamental equations describing rocking motion, first proposed by Housner \cite{Housner1963}, remain a cornerstone for understanding the response of rigid bodies under horizontal excitation.
Traditionally, research has focused on overturning as the primary failure condition, since preventing collapse is the main engineering objective. However, the nonlinear oscillations preceding overturning can exhibit complex behaviors, including transitions toward multiperiodic and chaotic regimes and the coexistence of competing attractors requiring deeper characterization.
In this context, impact dissipation, governed by the coefficient of restitution (COR), plays a critical role in determining the long-term dynamics of the system. Recent studies have emphasized that its treatment remains a key factor controlling the accuracy of rocking simulations \cite{RB_SP_NLD1}.
% Rocking blocks have been studied for decades due to their relevance in modeling and protecting structures such as masonry walls \cite{MasonryWalls2017}, statues \cite{AssymetricGeometry_4}, art pieces \cite{3DGeometry_2}, elevated water tanks \cite{WaterTank}, and other rigid-block-like objects requiring seismic protection \cite{RB_SP_NLD2}.
% The fundamental equations describing rocking motion, first proposed by Housner \cite{Housner1963}, remain a cornerstone for understanding the response of rigid bodies under horizontal excitation.
% Traditionally, research has focused on overturning as the primary failure condition, since preventing collapse is the main engineering objective. However, the nonlinear oscillations preceding overturning can exhibit complex behaviors, including transitions toward multiperiodic and chaotic regimes, as well as the coexistence of competing attractors that require deeper characterization.
% In this context, energy dissipation during impacts, governed by the coefficient of restitution (COR), plays a critical role in determining the long-term dynamics of the system. Recent studies have further emphasized that the treatment of impact dissipation remains one of the key factors controlling the accuracy of rocking simulations \cite{RB_SP_NLD1}.

Different expressions for the COR have been proposed to approximate the complex mechanics of rocking impacts.
The classical model introduced by Housner assumes conservation of angular momentum about the new pivot during collision, resulting in a restitution coefficient determined solely by block geometry.
Subsequent studies have proposed alternative interpretations of impact. Kalliontzis et al. \cite{Restitucion2016} suggested that the effective rotation center during impact may shift within the contact region instead of remaining fixed at the corner.
Ther and Kollár \cite{HousnerImprovement} showed experimentally that surface irregularities may produce multiple successive impacts rather than a single corner collision, reducing the effective energy loss. More recently, Mao et al. \cite{Restitucion2024} proposed an alternative theoretical formulation based on kinetic-energy redistribution, obtaining improved agreement with experimental measurements, particularly for low-slenderness blocks where discrepancies with the classical Housner model are most pronounced.
These modeling assumptions lead to distinct estimates of the energy dissipated during collision and may therefore produce different dynamical responses.

Although the nonlinear dynamics of rocking blocks has been extensively studied using the classical restitution coefficient, including analyses based on Lyapunov exponents \cite{Jiang2024}, bifurcation diagrams and basins of attraction \cite{Liu2021}, as well as global bifurcation mechanisms associated with heteroclinic structures and transitions to complex rocking responses \cite{lenciHeteroclinicBifurcationsOptimal2005}, comparatively little attention has been devoted to understanding how alternative restitution formulations influence the resulting dynamical landscape. In particular, while differences between experimental impact behavior and classical predictions have been reported, their influence on complex oscillatory regimes, attractor stability, and overturning susceptibility remains largely unexplored.

In this work, we compare the classical restitution model proposed by Housner with the formulation derived by Mao et al. and investigate how these impact assumptions affect the nonlinear dynamics of rocking blocks. Through bifurcation diagrams, Lyapunov exponents, and basins of attraction for different slenderness ratios, we characterize the dynamical consequences of adopting the alternative restitution formulation. The results show that the Mao et al. model promotes earlier transitions toward complex oscillatory behavior for low-slenderness blocks, while both formulations progressively converge as the block becomes more slender.

\section{Physical Model and Equations of Motion}

\begin{figure}[h]
\centerline{\includegraphics[width=0.8 \linewidth]{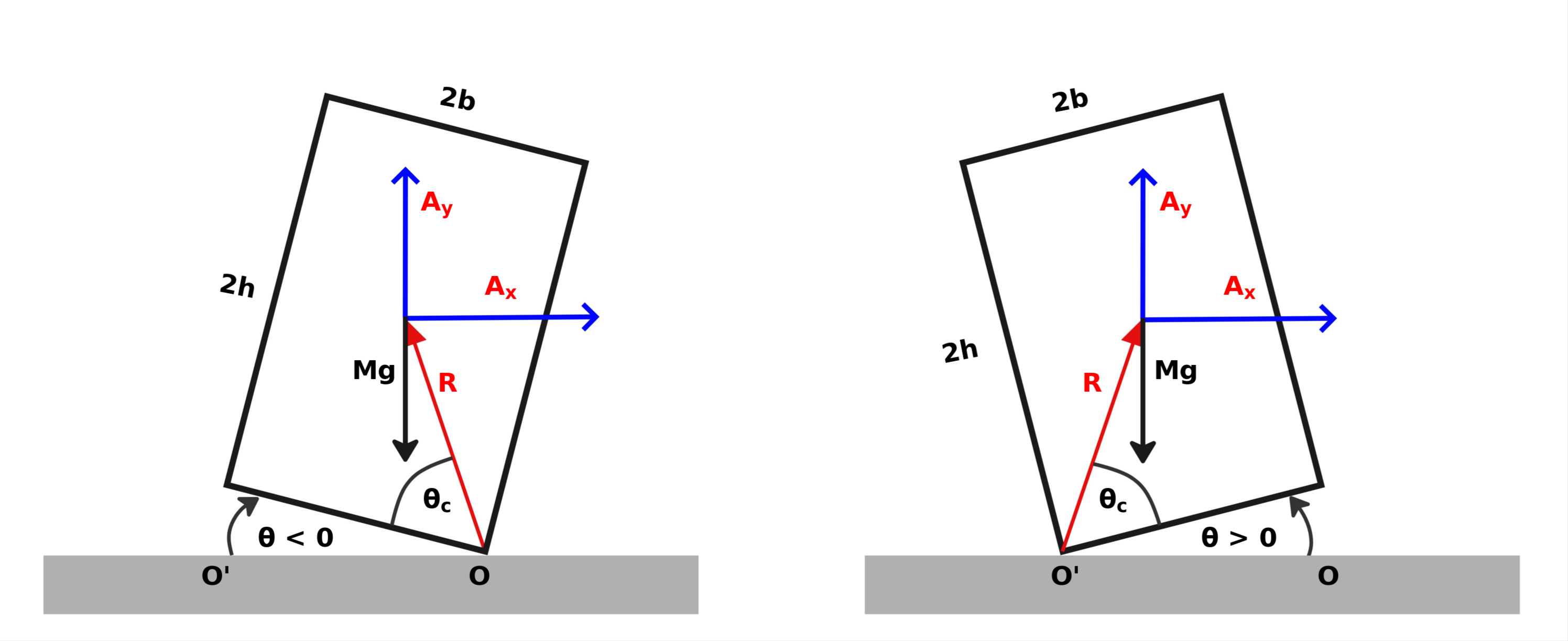}}
\caption{Diagram of the rocking block with pivots $O$ and $O'$.}
\label{fig:BlockDiagram}
\end{figure}

We consider the classical rigid rocking block introduced by Housner \cite{Housner1963}, consisting of a homogeneous rectangular body of mass $M$, width $2b$, and height $2h$, rocking without sliding on a rigid foundation. A schematic representation of the system and the active pivots is shown in Fig.~\ref{fig:BlockDiagram}.

The angular displacement of the block is described by the variable $\theta$, measured with respect to the horizontal ground. Positive and negative values of $\theta$ correspond to rocking about pivots $O'$ and $O$, respectively. The geometry is fully characterized by the slenderness ratio $h/b$, from which the critical angle $\theta_c=\arctan(h/b)$ is directly determined. The distance between the center of mass and the active pivot is given by $R=\sqrt{h^2+b^2}$.

The governing equations are obtained from the balance of moments about the active pivot and follow the classical rocking-block formulation originally proposed by Housner \cite{Housner1963} and subsequently employed in numerous studies of rocking dynamics such as \cite{Liu2021,Esther2024}. Using $R$ as the characteristic length and $\Omega_0=\sqrt{g/R}$ as the characteristic frequency, we introduce the dimensionless parameters

\begin{equation}
\omega=\frac{\Omega}{\Omega_0},
\qquad
\alpha=\frac{A}{R},
\qquad
\xi=\frac{c_b}{MR^2\Omega_0},
\end{equation}

where $\alpha$ and $\omega$ denote the amplitude and frequency of the harmonic base excitation, respectively, while $\xi$ represents the dimensionless viscous damping coefficient. The dimensionless angular velocity and acceleration are denoted by $\theta'$ and $\theta''$.

The forced rocking dynamics can then be written as

\begin{align}
\theta''-\frac{3}{4}f_p(t)\sin(\theta_c+\theta)
+\frac{3}{4}\cos(\theta_c+\theta)
+\frac{3}{4}\xi\theta'
&=0,
\qquad \theta>0,
\label{eq:dim_pos}
\\
\theta''-\frac{3}{4}f_p(t)\sin(\theta_c-\theta)
-\frac{3}{4}\cos(\theta_c-\theta)
+\frac{3}{4}\xi\theta'
&=0,
\qquad \theta<0,
\label{eq:dim_neg}
\end{align}

where the dimensionless forcing is given by $f_p(t)=\alpha\omega^2\sin(\omega t)$.

The physical parameters are fixed throughout this study as $M=2\times10^4,\mathrm{kg}$, $c_b=101,\mathrm{N,s,m}$, $g=9.81,\mathrm{m/s^2}$, and $R=3.764,\mathrm{m}$, while the slenderness ratio and forcing parameters are varied according to the analysis performed.

To obtain an autonomous representation of the periodically forced system, the harmonic excitation is incorporated through the standard oscillator embedding described in \cite{Osciladores2007,Strogatz}. Defining the state vector $z(t)=\left(\theta,\theta',r,s\right)^{\top}$ and the parameter vector $\beta=(\alpha,\omega,\xi,\theta_c)$, the resulting piecewise-autonomous system can be written as

\begin{equation}
z'(t)=f_{\pm}(z,\beta),
\end{equation}

with

\begin{equation}
f_{\pm}(z,\beta)=
\begin{pmatrix}
\theta' \\
\frac{3}{4}\left(
\alpha\omega^2 r\,\sin(\theta_c\pm\theta)
\mp\cos(\theta_c\pm\theta)
-\xi\theta'
\right) \\
r+\omega s-r(r^2+s^2) \\
s-\omega r-s(r^2+s^2)
\end{pmatrix},
\label{eq:vectorfield_full}
\end{equation}

where the upper sign corresponds to $\theta>0$ and the lower sign to $\theta<0$.

%%%%%%%%%%%%%%%%%%%%%%%%%%%%%%%%%%%%%%%%%%%%
\subsection{Impact Law and Restitution Models}\label{sec:RestitucionVel+}
When the rocking block crosses the vertical position ($\theta=0$), the active pivot changes instantaneously, triggering an impact with the supporting surface. In the classical rocking-block formulation, this collision occurs at time $t_k$. Denoting by $t^-$ and $t^+$ the instants immediately before and after impact, respectively, the angular position remains continuous, $\theta(t^+) = \theta(t^-)$, while the angular velocity changes according to the restitution coefficient $e$, such that $\theta'(t^+) = e\,\theta'(t^-)$.
All other state variables remain continuous across the impact. The coefficient $e$ represents the fraction of angular velocity preserved after collision and therefore controls the amount of energy dissipated with each pivot exchange.

\begin{figure}[htbp]
\centering
\includegraphics[width=0.6\textwidth]{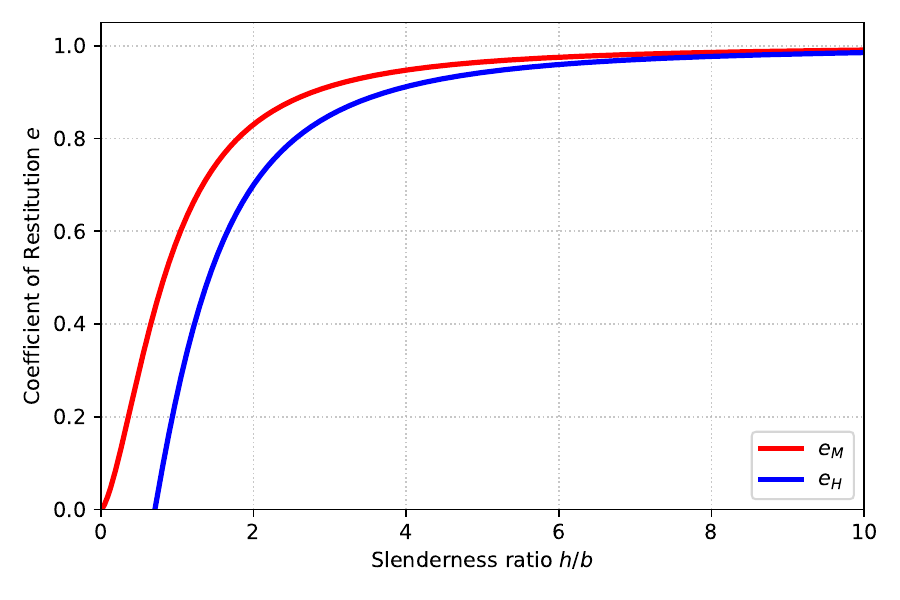}
\caption{Comparison between coefficients of restitution $e_H$ and $e_M$ for different critical angles. }
\label{fig:CompCoefs}
\end{figure}

\noindent The classical restitution coefficient derived by Housner \cite{Housner1963} is obtained from angular momentum conservation about the new pivot during impact. Under the assumptions of rigid body motion and no sliding, the coefficient depends only on the block geometry through the critical angle $\theta_c$,
\begin{equation}
e_H = 1 - \frac{3}{2}\cos^2(\theta_c).
\end{equation}
A more recent formulation proposed by Mao et al. \cite{Restitucion2024} derives the restitution coefficient from a vector-based analysis of the velocity field during impact, accounting for energy dissipation and elastic recovery at the contact interface. The resulting expression is
\begin{equation}
e_M^2 =
\frac{\tan(\theta_c)[\tan^2(\theta_c)-3]
+3\tan^4(\theta_c)\left(\frac{\pi}{2}-\theta_c\right)
+3\theta_c}{4\tan(\theta_c)\sec^2(\theta_c)}.
\label{eq:mao_cor}
\end{equation}
Experimental comparisons reported in \cite{Restitucion2024} indicate that the proposal by Mao et al. provides restitution values closer to measured impact data and generally predicts lower energy dissipation than the classical Housner expression. 
Figure~\ref{fig:CompCoefs} compares both coefficients as a function of the critical angle $\theta_c=\tan^{-1}(h/b)$. The comparison reveals a qualitative difference between the models for low slenderness ratios, while both formulations converge  to $e\rightarrow1$ as $\theta_c$ increases.

To quantify the discrepancy between both models, we use the symmetric percentage difference \cite{SymDifference}, which provides a normalized comparison without selecting either formulation as reference,

\begin{equation}
\Delta e \% =
\frac{|e_M - e_H|}{0.5(e_M + e_H)}\times 100\%.
\end{equation}

% It highlights the magnitude of the discrepancy in a consistent manner across different parameter values, facilitating comparison between the restitution formulations.

\subsection{Numerical Characterization of the Dynamics}
\label{sec:Lyapunov}

The dynamical response of the rocking block is characterized using bifurcation diagrams, maximum Lyapunov exponents, and basins of attraction. Together, these tools provide complementary information regarding the long-term behavior, stability, and organization of the phase space.

Bifurcation diagrams are employed to identify qualitative changes in the asymptotic response as a control parameter varies, revealing transitions between periodic, multiperiodic, and chaotic regimes \cite{RefBifurca2,LimitationHopf1,Strogatz,RefBifurca1}.

The maximum Lyapunov exponent $\lambda_{\max}$ is computed using the Benettin--Wolf algorithm with QR re-orthonormalization \cite{Lyapunov1992,LyapunovIntro1}. Between impacts, perturbations evolve according to the variational equations, while discontinuities are incorporated through saltation matrices for nonsmooth systems \cite{LyapunovDiscontinous}. Following \cite{LyapunovRB}, the saltation matrix is adapted to the present four-dimensional vector fields (Eq.~\eqref{eq:vectorfield_full}) and their associated Jacobians. Positive values of $\lambda_{\max}$ indicate chaotic dynamics, whereas negative values correspond to stable periodic responses.

% The maximum Lyapunov exponent $\lambda_{\max}$ is computed using the Benettin--Wolf algorithm with QR re-orthonormalization \cite{Lyapunov1992,LyapunovIntro1}. Between impacts, perturbations evolve according to the variational equations of the continuous system, while impact-induced discontinuities are incorporated through saltation matrices for nonsmooth systems \cite{LyapunovDiscontinous}. Following the framework proposed in \cite{LyapunovRB}, the saltation matrix is adapted here to the present four-dimensional vector fields (Eq.~\eqref{eq:vectorfield_full}) and their associated Jacobians. Positive values of $\lambda_{\max}$ indicate chaotic dynamics, whereas negative values correspond to stable periodic responses.

Basins of attraction are used to characterize the dependence of the long-term dynamics on initial conditions by partitioning the state space into regions associated with distinct asymptotic behaviors \cite{Strogatz,SpringsBasins,AssymetricGeometry}. This representation allows the identification of damped motion, sustained rocking oscillations, multiperiodic responses, and overturning, defined here as trajectories satisfying $|\theta| \geq \pi/2$.

Our simulations were implemented in \texttt{Python 3} using the \texttt{SciPy} library. The nonlinear hybrid dynamical system was integrated using the \texttt{solve\_ivp} routine from \texttt{scipy.integrate} with the implicit \texttt{BDF} scheme \cite{SciPyBDF}. The numerical tolerances were fixed at \texttt{rtol}$=10^{-6}$ and \texttt{atol}$=10^{-9}$.

Event detection was used to monitor the condition $\theta=0$, corresponding to the impact event and pivot exchange, as well as the overturn condition $|\theta|=\pi/2$. When an impact is detected, the integration is stopped at the event time and the angular velocity is updated according to the restitution law described previously.

To assess the influence of the restitution model on the global dynamics, we compare the coefficients $e_H(\cdot)$ and $e_M(\cdot)$ under identical geometric, excitation, and initial conditions. To ensure that differences in the dynamical indicators arise solely from the impact model, the initial condition for the bifurcation and Lyapunov analyses is

\begin{equation}
\mathbf{z}_0 =
\left(
-0.2254,\,
-0.1597,\,
-0.7778,\,
0.6284
\right)^{\top}.
\label{eq:initial_condition}
\end{equation}
Two slenderness ratios are considered, $h/b \in \{1,2\}$. For each geometry and restitution model, the control parameter $\alpha$ is swept within the intervals reported in Table~\ref{tab:SystemParameters}, using a uniform step $\Delta \alpha = 0.001$. In all cases the excitation frequency $\omega$ is kept constant and selected to ensure an active dynamical response for the corresponding geometry. The remaining physical parameters are fixed across all simulations.
\begin{table}[htbp]
\caption{Simulation parameters.} \label{tab:SystemParameters}
{\tabcolsep8pt
\begin{tabular}{lcc}
\toprule
 & $h/b=1$ & $h/b=2$ \\
\midrule
Excitation frequency $\omega$ 
& $1.30$ & $1.60$ \\
Range $\alpha$ (Housner) 
& $[0.50,\,1.00]$ 
& $[0.10,\,0.50]$  \\
Range $\alpha$ (Mao et al.) 
& $[0.30,\,0.80]$ 
& $[0.10,\,0.50]$ \\
$e_H$ & $0.250$ & $0.700$ \\
$e_M$ & $0.582$ & $0.830$ \\
$\Delta e\ (\%)$ & $79.85$ & $17.03$ \\
\botrule
\end{tabular}}
\end{table}

For each value of the control parameter $\alpha$, the system is integrated over $500$ excitation periods $T = 2\pi/\omega$. To eliminate transient effects, only the final $100$ periods, corresponding to $t \in [400T,\,500T]$, are retained and taken as representatives of the asymptotic behavior, following \cite{Liu2021}.

Bifurcation diagrams are constructed from Poincaré section points in the $(\alpha,\theta')$ plane, while the maximum Lyapunov exponent is evaluated as a function of $\alpha$. In both cases, only data from the asymptotic window are used, ensuring that the observed structures correspond to the established long-term dynamics.

The maximum Lyapunov exponent $\lambda_{\max}$ is computed during the numerical integration using the Benettin--Wolf algorithm described previously. The dynamical system and its associated variational equations are integrated simultaneously, with the perturbation vector initialized with norm $10^{-8}$ and periodically renormalized using QR re-orthonormalization. At each impact event, the tangent dynamics is updated using the saltation matrix for nonsmooth systems, ensuring the correct propagation of perturbations across the discontinuity. The exponent $\lambda_{\max}$ is estimated as the average logarithmic growth rate of the perturbation vector over the asymptotic window.

Basins of attraction are constructed using representative values of $\alpha$ identified in the bifurcation diagrams for which sustained oscillations occur. For each the initial condition space $(\theta_0,\theta'_0) \in[-1,1]\times[-1,1]$ is discretized using a uniform $100\times100$ grid. Each initial condition is integrated until the asymptotic behavior is reached and subsequently classified according to the dynamical criteria defined previously. 
To isolate the effect of the restitution model, the initial phase of the excitation is fixed at $r_0 = 0.0$ and $s_0 = 1.0$, so that variations in the basins arise solely from the block initial conditions and the adopted impact model.

\section{Results}
\label{sec:ResultadosEstDinamica}
\begin{figure}[h]
\centering
\includegraphics[width=0.8\columnwidth]{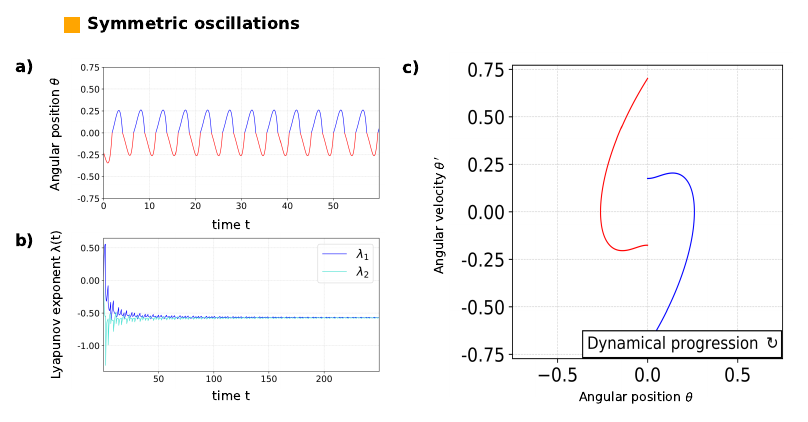}
\caption{
Characterization of symmetric motion. 
Panel a): time series $\theta(t)$.
Panel b): maximum Lyapunov exponent time series $\lambda (t)$.
Panel c): phase space $(\theta,\theta')$. Parameters used: 
$\alpha = 0.6$, $\omega = 1.3$, $h/b = 1$. Initial conditions \eqref{eq:initial_condition}.}
\label{fig:ReferenceCase}
\end{figure}

\begin{figure}[h]
\centering
\includegraphics[width=0.8\columnwidth]{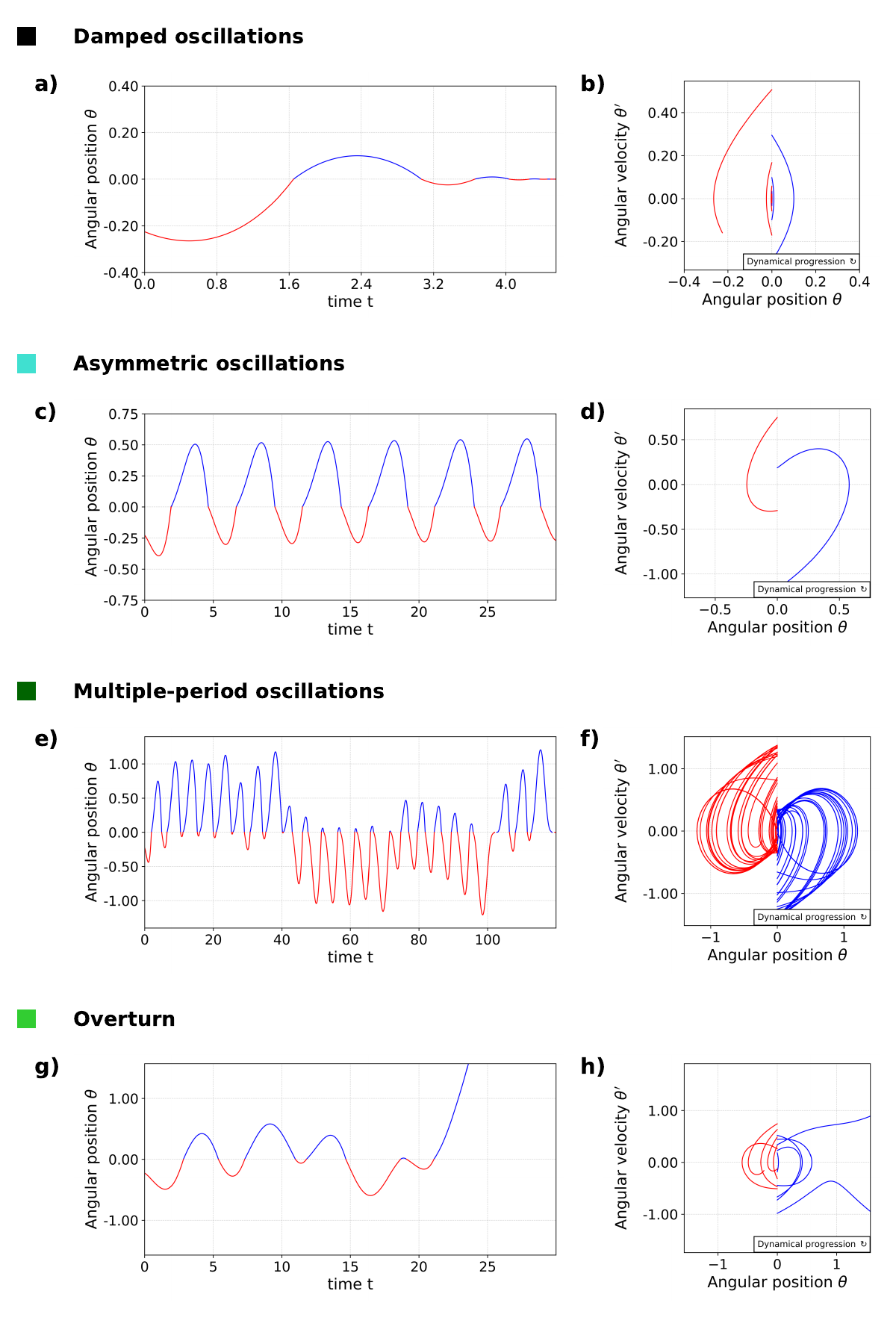}
\caption{
Characterization of dynamical behaviors. 
Left and right columns display time series $\theta(t)$ and phase space $(\theta,\theta')$, respectively. 
All scenarios use the initial conditions \eqref{eq:initial_condition} and the Housner formulation, except for damped oscillations. 
The parameters $(\alpha,\omega,h/b)$ associated with each behavior are indicated alongside each row: 
Row 1: damped oscillations $(0.15,1.0,1)$; 
Row 2: asymmetric oscillations $(0.8,1.3,1)$;
Row 3: multiple-period oscillations $(0.969,1.3,1)$; 
Row 4: overturn $(0.448,1.3,2)$.
}
\label{fig:BehaviorMosaic1}
\end{figure}

To establish a reference for the dynamical characterization, Fig.~\ref{fig:ReferenceCase} presents a representative symmetric oscillatory response. The time series $\theta(t)$ exhibits alternating amplitudes of equal magnitude, while the phase portrait forms an orbit symmetric with respect to the origin. The evolution of the maximum Lyapunov exponent $\lambda(t)$ provides the corresponding stability characterization. Building on this reference case, the remaining dynamical behaviors observed in the system are summarized in Fig.~\ref{fig:BehaviorMosaic1} through their time series and phase portraits.

% Building on this reference case, Fig.~\ref{fig:BehaviorMosaic1} summarizes the main dynamical behaviors observed in the system through time series and phase portraits. 

\FloatBarrier

\begin{figure}[htbp]
\centering
\includegraphics[width=0.7\columnwidth]{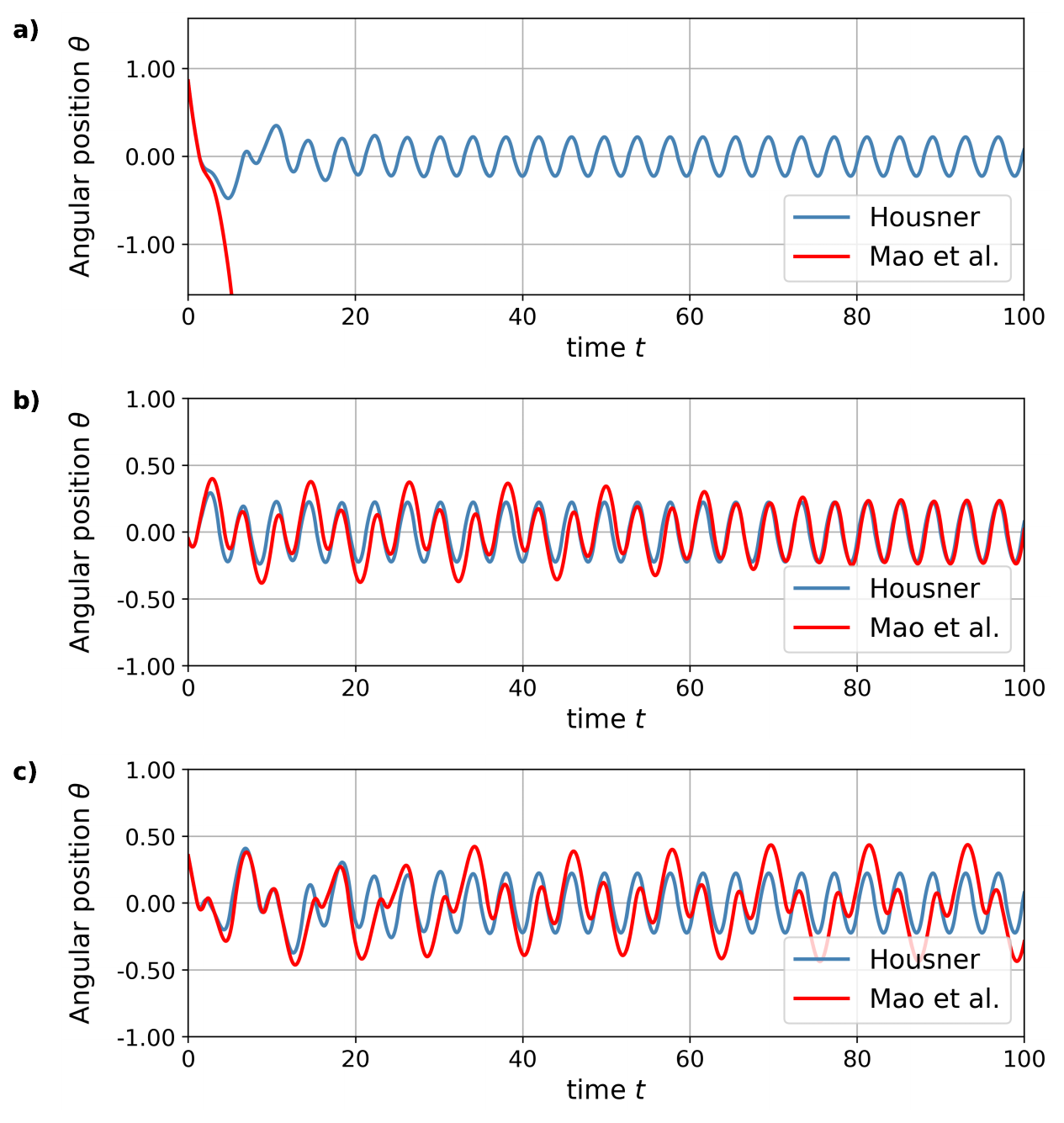}
\caption{
Comparison of time series $\theta(t)$ under identical parameters using the Housner model and the formulation of Mao et al. 
a) divergence toward overturning, 
b) symmetric oscillations, and 
c) differences in amplitude and multiperiodic structure. 
All cases use $\alpha = 0.2$, $\omega = 1.6$, and $h/b = 2$, with $r_0 = 0$ and $s_0 = 1$. 
The initial conditions $(\theta_0,\theta'_0)$ for cases a)--c) are 
$(0.8586,-0.8182)$, $(-0.0505,-0.2121)$, and $(0.3535,-0.3333)$, respectively.
}
\label{fig:TimeSeriesComparison}
\end{figure}

\FloatBarrier
These behaviors define the classification framework used in the following analysis and serve as a basis for direct comparison between restitution models under identical conditions, as illustrated in Fig.~\ref{fig:TimeSeriesComparison}. 

% \FloatBarrier

The responses produced by the Housner and Mao et al. formulations exhibit clear qualitative differences, including divergence toward overturning, variations in oscillation amplitude and multiperiodic structure, and distinct convergence patterns toward bounded oscillatory states. Since all cases share identical excitation parameters and initial conditions, these differences arise solely from the adopted restitution model. The results therefore illustrate how changes in impact dissipation modify the cycle-by-cycle energy transfer and can ultimately alter the long-term dynamical regime predicted for the same physical configuration.

\begin{figure}[h]
    \centering
    % Primera imagen (el mosaico de bifurcación)
    \includegraphics[width=0.91\columnwidth, trim=0 0.7cm 0 0, clip]{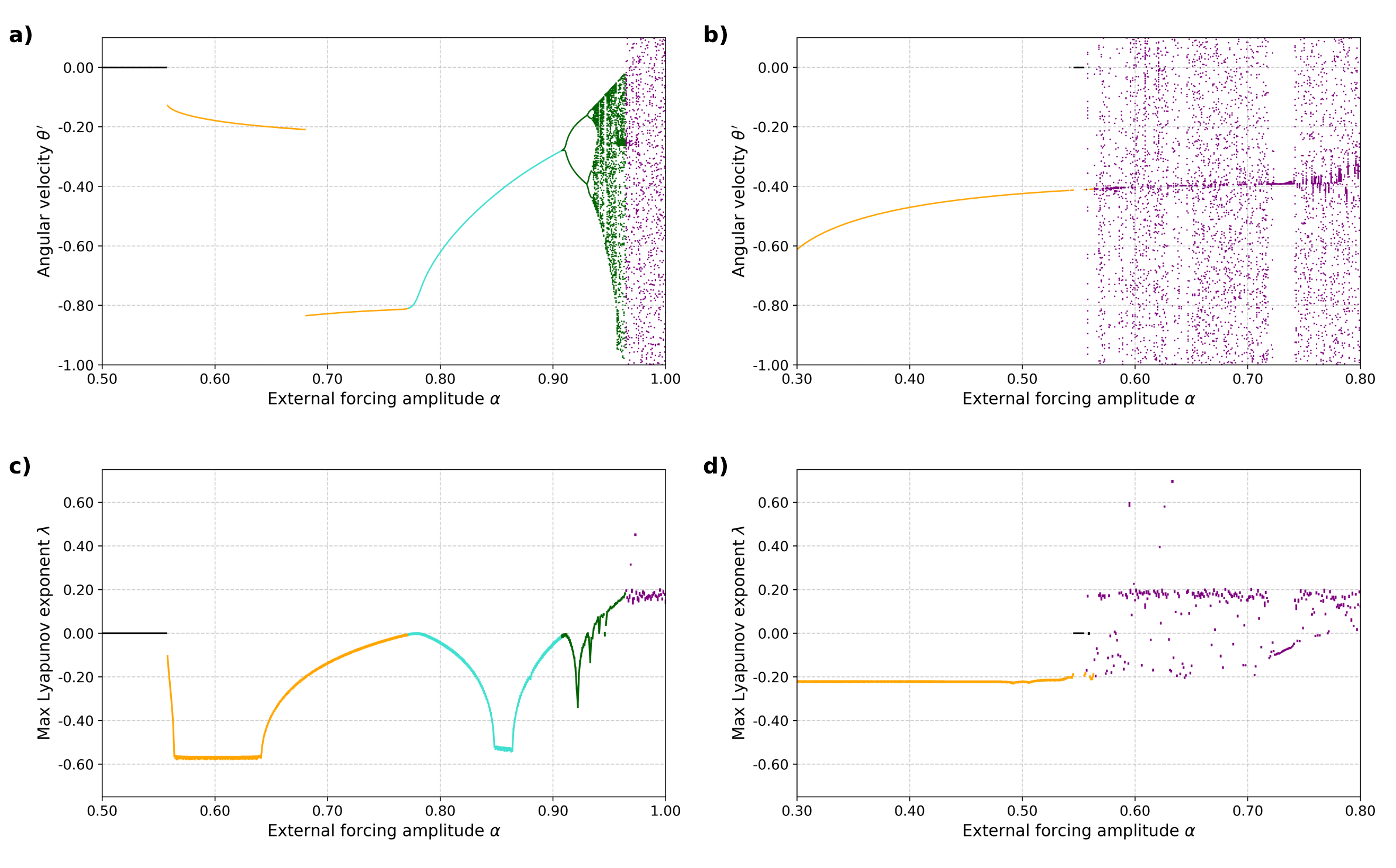}
    
    \vspace{0.2cm} % Espacio vertical opcional entre la imagen y la 
    \includegraphics[width=0.9\columnwidth, trim=0 0.5cm 0 0.5cm, clip]{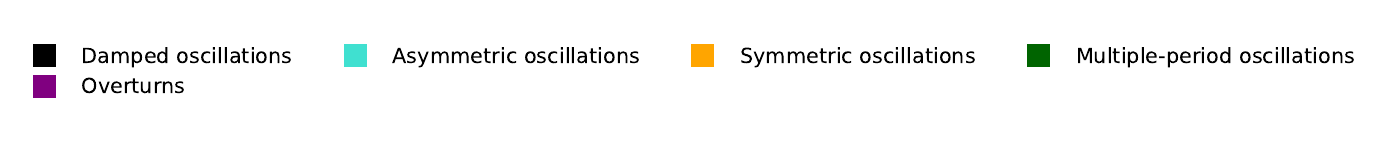}
    
    \caption{Bifurcation diagrams and maximum Lyapunov exponents for $h/b=1$. a) Bifurcation diagram for the Housner restitution model. b) Bifurcation diagram for the Mao et al. restitution model. c) Maximum Lyapunov exponent $\lambda_{\max}$ for the Housner model. d) Maximum Lyapunov exponent $\lambda_{\max}$ for the Mao model. Colors indicate the dynamical behaviors defined in Figures~\ref{fig:ReferenceCase} and \ref{fig:BehaviorMosaic1}, according to the classification labels shown in the legend. 
    } 
\label{fig:Mosaic_hb1}
\end{figure}

Building on the identified dynamical behaviors, Fig.~\ref{fig:Mosaic_hb1} summarizes the system response as the control parameter $\alpha$ varies for $h/b=1$. The Housner and Mao et al. restitution models are shown side by side (left and right columns), with bifurcation diagrams in the upper panels and maximum Lyapunov exponents in the lower panels. This combined representation enables a joint characterization of the dynamical transitions induced by the forcing amplitude. For both restitution models, low forcing amplitudes are associated with damped or simple periodic responses, characterized by single-branch structures in the bifurcation diagrams and negative values of $\lambda_{\max}$, indicating stable non-chaotic motion. As $\alpha$ increases, additional periodic branches emerge, revealing period-multiplication and asymmetric oscillatory regimes. In these regions, $\lambda_{\max}$ may remain negative, showing that multi-period responses can still correspond to stable periodic attractors rather than chaotic dynamics.

The two formulations nevertheless produce markedly different global responses. For the Housner model, the transition toward irregular behavior occurs progressively, with alternating windows of multi-periodicity and positive Lyapunov exponents before the onset of overturning trajectories. In contrast, the Mao et al. model shifts the onset of complex regimes toward lower values of $\alpha$, exhibiting an earlier and more abrupt loss of regular periodic motion, with broad parameter intervals displaying scattered bifurcation structures and predominantly positive $\lambda_{\max}$ values. Since both formulations are evaluated under identical excitation and geometric conditions, these differences arise solely from the adopted restitution law. The results therefore indicate that the impact model influences not only the amount of energy dissipated during collisions, but also the location of dynamical transitions and the extent of parameter regions associated with complex responses. This effect is particularly pronounced for low-slenderness blocks and is consistent with the reduced effective dissipation predicted by the Mao et al. formulation. For higher slenderness ratios, the differences between both models become less pronounced, with similar bifurcation structures and comparable distributions of $\lambda_{\max}$, consistent with the convergence of the restitution coefficient as $h/b$ increases. A magnified view of selected regions of the Housner bifurcation structure is provided in Appendix~\ref{fig:BifurcationAppendix}, while the corresponding diagrams and basins for $h/b=2$ are included in Appendix~\ref{fig:Mosaic_hb23} and Appendix~\ref{fig:Basins2}, respectively.

For trajectories converging to periodic attractors, $\lambda_{\max}$ yields a well-defined value for each $\alpha$. Some trajectories classified as overturning also exhibit negative values of $\lambda_{\max}$, corresponding to cases where the motion re-enters the bounded domain after the overturn event and converges. These responses are nevertheless classified as overturning, since the escape condition is met at least once. This criterion also accounts for the reduced scattering observed in the bifurcation diagrams for certain values of $\alpha$.

\begin{figure}[h!]
\centering
\includegraphics[width=0.9\columnwidth]{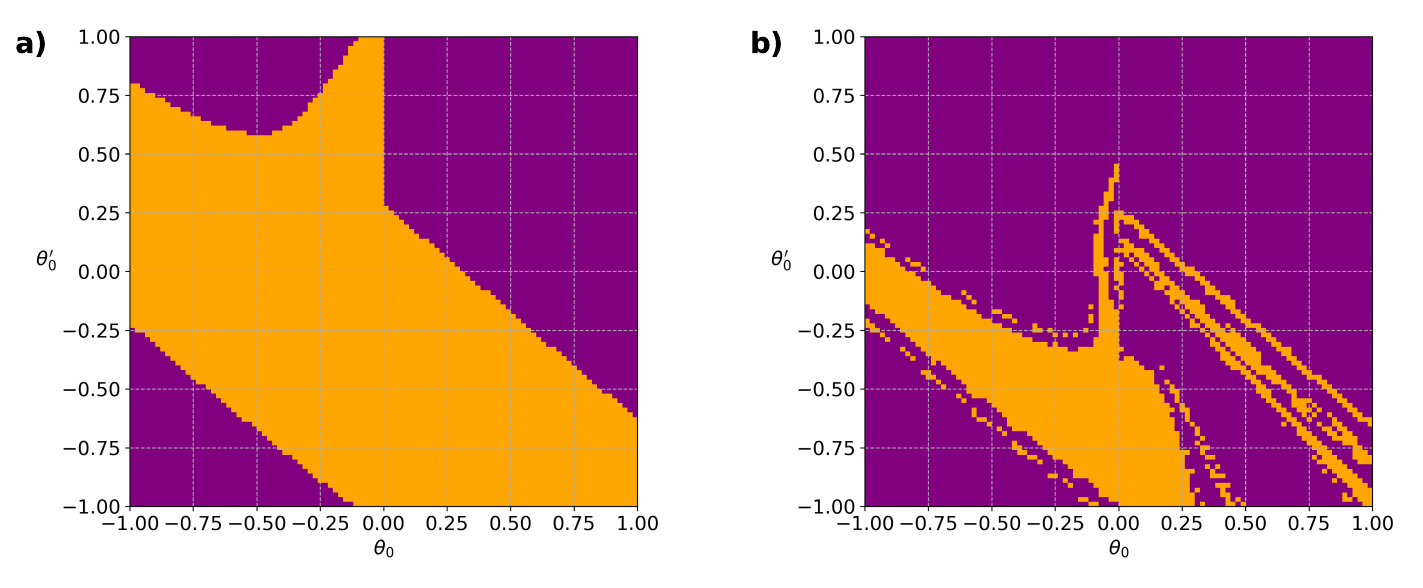}
    \includegraphics[width=0.95\columnwidth, trim=0 1cm 0 1cm, clip]{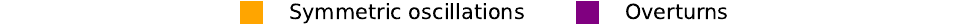}
\caption{Basins of attraction in the $(\theta_0,\theta'_0)$ plane for $\alpha = 0.66$, $\omega = 1.3$.
a) Housner restitution model.
b) Mao et al. restitution model.
Colors correspond to the labels shown in the legend.}
\label{fig:Basins}
\end{figure}

Figure~\ref{fig:Basins} shows the basins of attraction in the $(\theta_0,\theta'_0)$ plane for the restitution models considered. The Housner model exhibits a comparatively broad and compact basin associated with symmetric oscillations, with relatively smooth boundaries separating bounded responses from overturning trajectories. This indicates a higher degree of robustness with respect to perturbations in the initial state. In contrast, the Mao et al. model displays a substantially narrower oscillatory basin, organized in thin bands with more irregular boundaries embedded within a dominant overturning region. Such fragmentation reveals an increased sensitivity to initial conditions and a reduced domain of stable bounded motion. Consequently, identical perturbations of the initial state may lead to qualitatively different outcomes depending solely on the adopted restitution model, highlighting the influence of impact modeling on the global accessibility of bounded rocking responses.

Similarly to the basins reported in \cite{Liu2021, SpringsBasins, AssymetricGeometry}, the structure of the phase space exhibits a dominant partition between overturning domains and sustained oscillations, further subdivided into symmetric, weak, and multiperiodic behaviors. While this global organization is consistent with previous studies, the finer classification adopted here allows the redistribution of initial conditions among these behaviors to be quantified, as summarized in Table~\ref{tab:BasinsCount}.

\begin{table}[htbp]
\caption{Percentage distribution of grouped dynamical behaviors for $10\,000$ initial conditions.}\label{tab:BasinsCount}
{\tabcolsep8pt
\begin{tabular}{lcccc}
\toprule
 & \multicolumn{2}{c}{$h/b = 1$}
 & \multicolumn{2}{c}{$h/b = 2$} \\
\midrule
Behavior
 & H (\%) & M (\%)
 & H (\%) & M (\%)\\
\midrule
Overturning      & 44.45 & 80.36 & 80.33 & 91.02\\
Symmetric osc.   & 55.55 & 19.64 & 19.67 & 6.68   \\
Multiple period osc.    & 0.00  & 0.00  & 0.00  & 2.30   \\
\botrule
\end{tabular}}
\end{table}

\noindent The agreement between models was computed point-wise over the grid of $10\,000$ initial conditions, counting only the cases in which both formulations predict the same dynamical behavior under identical excitation parameters. For $h/b = 1$, the overall agreement is $64.09\%$, whereas for $h/b = 2$ it increases to $86.93\%$. Thus, more than one third of the initial conditions for the low-slenderness block lead to different dynamical outcomes depending solely on the adopted restitution model. For $h/b = 1$, the Mao et al. formulation increases the proportion of overturning events and reduces the oscillatory domains. As $h/b$ increases, the relative proportions between the models become closer and the organization of the phase space progressively converges.

Beyond the local transitions identified through bifurcation diagrams and Lyapunov exponents, the basin analysis reveals that the restitution law also modifies the global accessibility of bounded motions in phase space, affecting the set of initial conditions from which safe oscillatory responses can be attained. For $h/b=1$, the Mao et al. formulation substantially contracts the basin associated with regular symmetric oscillations, increasing the prevalence of overturning outcomes under perturbations of the initial state. This indicates that changes in impact dissipation affect not only the stability thresholds, but also the practical robustness of safe rocking responses.
For $h/b=2$, the differences become more subtle in terms of global agreement, yet the Mao model still produces a richer internal organization of the bounded region through the appearance of multiperiodic attractors. This suggests that as slenderness increases, the influence of restitution modeling shifts from strongly altering basin size toward reshaping the diversity of admissible long-term oscillatory states.

\section{Conclusions}

This work investigated the influence of restitution modeling on the nonlinear dynamics of rocking blocks subjected to harmonic excitation. The comparison between the classical coefficient proposed by Housner and the alternative expression derived by Mao et al. reveals that the adopted impact model plays a significant role in shaping the global dynamical structure of the system.

The results show that the Mao et al. model predicts higher restitution values, leading to reduced impact dissipation and promoting earlier transitions toward complex oscillatory behavior. These differences are reflected in the bifurcation diagrams, the distributions of the maximum Lyapunov exponent, and the organization of the basins of attraction. In particular, for low-slenderness blocks the Mao et al. formulation significantly reduces the domain of regular bounded oscillations and increases the prevalence of overturning responses, whereas for higher slenderness ratios it additionally favors the emergence of multiperiodic bounded motions. Under identical excitation conditions, the two restitution models may therefore predict qualitatively different long-term responses, highlighting the sensitivity of rocking dynamics to the adopted impact law.

Across the explored parameter space, large regions exhibit well-defined dynamical regimes characterized by a single dominant behavior, such as periodic rocking, multiperiodic oscillations, or overturning. However, consistent with the strongly nonlinear and hybrid nature of the rocking system, regions also appear in which different dynamical responses coexist in close proximity. In these mixed zones, small variations in parameters or initial conditions may lead to qualitatively different long-term outcomes, reflecting the presence of competing attractors and sensitive dependence on initial conditions typical of impact-driven nonlinear systems.

Beyond the local transitions identified through bifurcation diagrams and Lyapunov exponents, the basin analysis reveals that restitution modeling also affects the global accessibility of bounded motions in phase space. For $h/b=1$, the agreement between both models is only $64.09\%$, indicating that more than one third of the evaluated initial conditions lead to different dynamical outcomes depending solely on the adopted restitution formulation. As the slenderness ratio increases, the corresponding dynamical indicators progressively converge, consistent with the reduced discrepancy between the restitution coefficients.

Overall, the results show that the coefficient of restitution should not be regarded merely as a local impact parameter, but as a modeling assumption capable of altering attractor organization, stability boundaries, and overturning susceptibility. More broadly, the present study provides a dynamical interpretation of how alternative restitution formulations, introduced to address discrepancies between classical predictions and experimentally observed impact behavior, influence the long-term response of rocking systems. These findings contribute to a more comprehensive understanding of rocking stability by clarifying how impact dissipation interacts with geometric properties to determine the observed dynamical behavior.

Although the restitution formulation adopted in this work provides a closer approximation to experimentally observed impact behavior, the model still relies on the idealized assumptions of rigid-body dynamics and high-friction conditions that prevent sliding. In real structural systems, contact interactions may involve local deformation, frictional slip, and viscoelastic dissipation mechanisms that are not captured by the simplified restitution framework considered here. Consequently, future developments aimed at incorporating material properties and contact mechanics into the estimation of the restitution coefficient may provide a more realistic representation of impact dissipation and further refine the predicted stability boundaries of rocking systems.

\section*{Declarations}

\begin{itemize}

\item \textbf{Funding}  
The authors received no specific funding for this work.

\item \textbf{Conflict of interest}  
The authors declare that they have no conflict of interest.

\item \textbf{Ethics approval and consent to participate}  
Not applicable.

\item \textbf{Consent for publication}  
Not applicable.

\item \textbf{Data availability}  
The datasets generated during the current study are available from the corresponding author upon reasonable request.

\item \textbf{Materials availability}  
Not applicable.

\item \textbf{Code availability}  
The numerical code used to generate the results of this study is available from the corresponding author upon reasonable request.

\item \textbf{Author contributions}

Fernando Gaibor E.: Conceptualization, numerical implementation, computational experiments, data analysis, visualization, and writing of the original manuscript.

Alexander López: Supervision, methodological discussion, interpretation of results, manuscript review, and editing.

Esther D. Gutiérrez: Conceptual guidance on rocking dynamics, supervision, validation of the methodology, manuscript review, and editing. Esther D. Gutiérrez also provided the initial computational framework upon which the present study was developed.

All authors reviewed and approved the final manuscript.

\end{itemize}

%%===================================================%%
%% For presentation purpose, we have included        %%
%% \bigskip command. Please ignore this.             %%
%%===================================================%%
%\bigskip
%\begin{flushleft}%
%Editorial Policies for:

%\bigskip\noindent
%Springer journals and proceedings: \url{https://www.springer.com/gp/editorial-policies}

%\bigskip\noindent
%Nature Portfolio journals: %\url{https://www.nature.com/nature-research/editorial-policies}

%\bigskip\noindent
%\textit{Scientific Reports}: %\url{https://www.nature.com/srep/journal-policies/editorial-policies}

%\bigskip\noindent
%BMC journals: %\url{https://www.biomedcentral.com/getpublished/editorial-policies}
%\end{flushleft}

\begin{appendices}

\section{Detailed bifurcation structure for the Housner model ($h/b=1$)}\label{ap:Appendix1}

\begin{figure}[htbp]
\centering
\includegraphics[width=0.7\columnwidth]{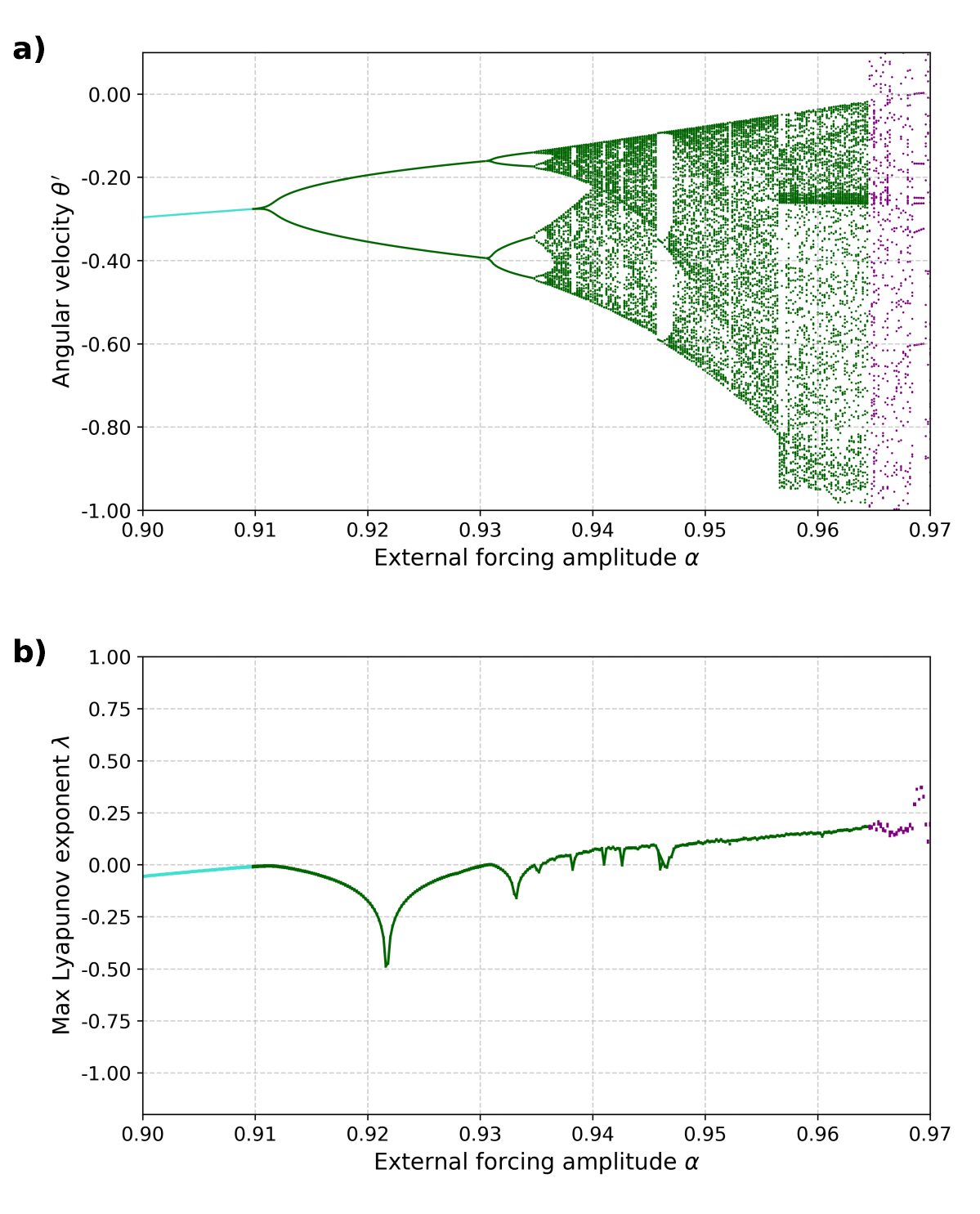}
\includegraphics[width=0.7\columnwidth, trim=0 1cm 0 1cm, clip]{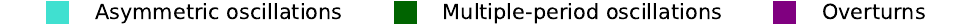}
\caption{
Detailed view of the Housner bifurcation structure for $h/b=1$.
a) Bifurcation diagram in the $(\alpha,\theta')$ plane.
b) Maximum Lyapunov exponent $\lambda_{\max}$.
Colors indicate the dynamical behaviors.}
\label{fig:BifurcationAppendix}
\end{figure}

Figure~\ref{fig:BifurcationAppendix} shows a magnified view of selected regions of the Housner bifurcation structure for $h/b=1$. As the forcing amplitude increases, the response evolves from asymmetric periodic oscillations to higher-period motions through successive bifurcations. The associated Lyapunov exponent remains negative during the first multiperiodic windows, confirming the persistence of stable bounded periodic attractors, and becomes positive only after further loss of stability, indicating the onset of chaotic dynamics. The non-monotonic variations of $\lambda_{\max}$ across these transitions reflect the hybrid impact nature of the rocking system, where alternating phases of contraction and expansion arise as the orbital structure changes.

\FloatBarrier
% \vspace{0.8cm}
\section{Bifurcation and Lyapunov exponents for $h/b=2$ block}\label{ap:Appendix2}
\begin{figure}[h]
\centering
    \includegraphics[width=0.9\columnwidth, trim=0 0.8cm 0 0.8cm, clip]{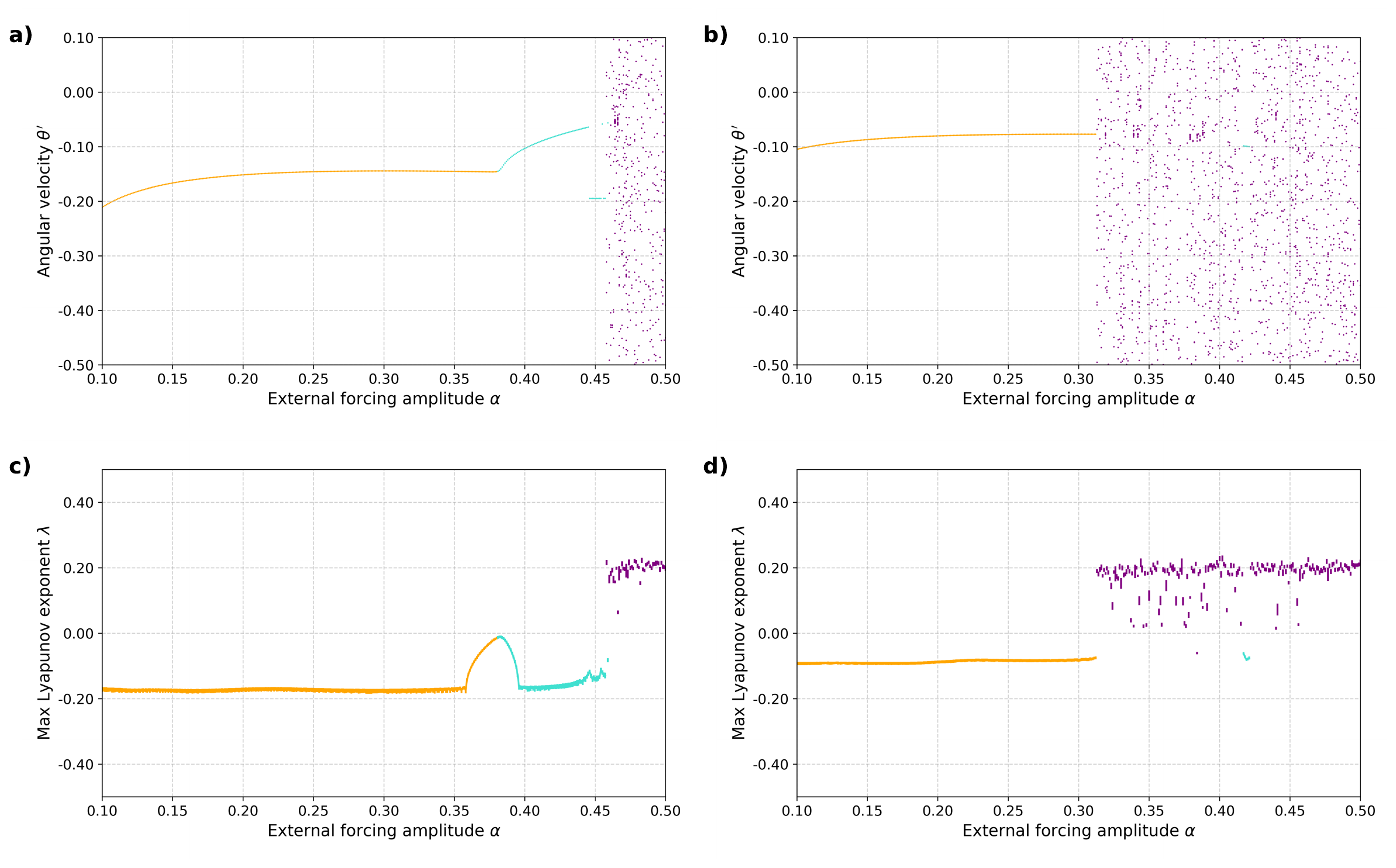}
    \includegraphics[width=0.8\columnwidth, trim=0 0.8cm 0 0.8cm, clip]{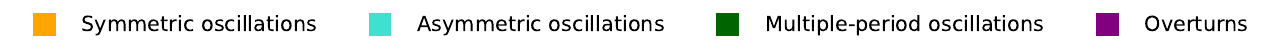}
\caption{Bifurcation diagrams and maximum Lyapunov exponents for $h/b = 2$.
Left column: Housner restitution model.
Right column: Mao et al. restitution model.
Rows 1 and 3: bifurcation diagrams.
Rows 2 and 4: maximum Lyapunov exponent $\lambda_{\max}$.}
\label{fig:Mosaic_hb23}
\end{figure}

Figure~\ref{fig:Mosaic_hb23} presents the bifurcation diagrams and maximum Lyapunov exponents for $h/b=2$. Compared with the lower slenderness case, the transition from oscillatory regimes to overturning occurs more abruptly, reflecting the increased geometric instability. A localized increase in $\lambda_{\max}$ is observed near the transition between symmetric and asymmetric oscillations, forming a small hump similar to that identified for $h/b=1$ and preceding the onset of more complex dynamics.

\FloatBarrier

% {\small
\section{Basins of attraction for $h/b=2$ block}\label{ap:Appendix3}
\begin{figure}[htb]
\centering
\includegraphics[width=0.85\columnwidth]{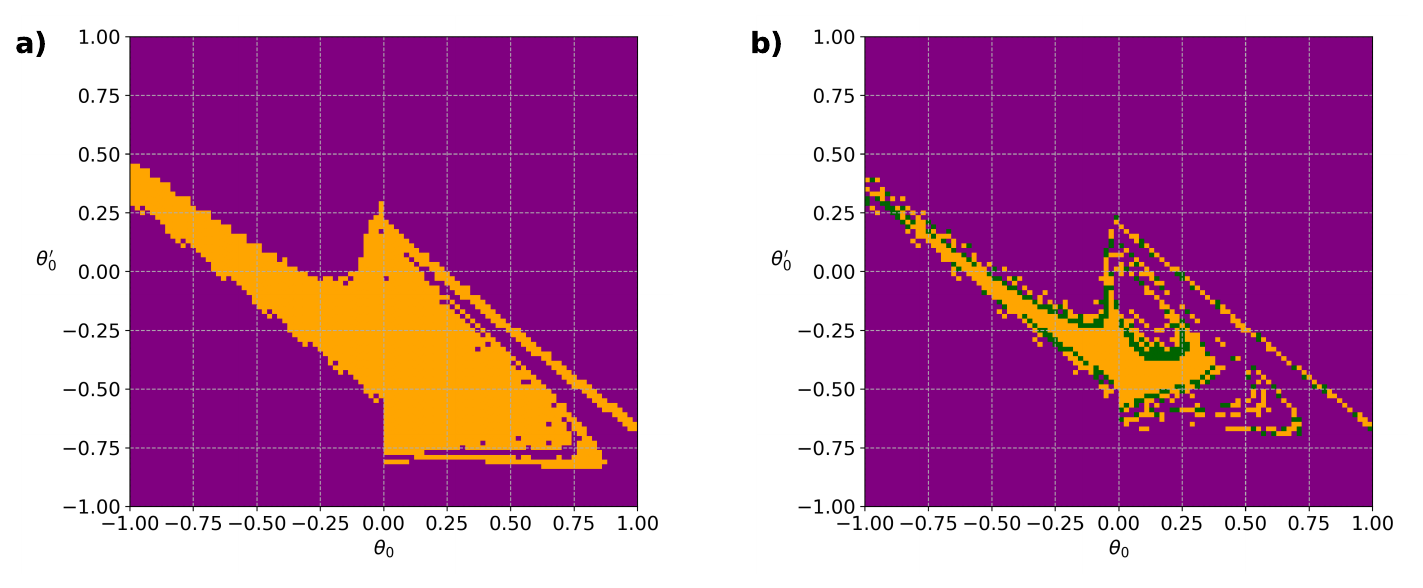}
    \includegraphics[width=0.8\columnwidth, trim=0 1cm 0 1cm, clip]{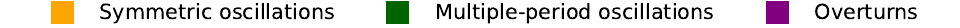}
\caption{Basins of attraction in the $(\theta_0,\theta'_0)$ plane for $\alpha = 0.2$ and $\omega = 1.6$.
a) Housner model.
b) Mao et al. model.
Colors indicate the dynamical behaviors.}
\label{fig:Basins2}
\end{figure}

% }
Figure~\ref{fig:Basins2} shows the basins of attraction for $h/b=2$. For the Housner model, the phase space remains divided between symmetric oscillations and overturning trajectories, with regular basin boundaries. In contrast, the Mao et al. formulation produces intricate geometries, with interwoven regions of different asymptotic behaviors and visible multiperiodic regions. This organization indicates increased coexistence of bounded responses, consistent with reduced impact dissipation favoring persistent dynamics.

\FloatBarrier
\end{appendices}

\FloatBarrier

\newpage

% \begin{appendices}

% \section{Section title of first appendix}\label{secA1}

% An appendix contains supplementary information that is not an essential part of the text itself but which may be helpful in providing a more comprehensive understanding of the research problem or it is information that is too cumbersome to be included in the body of the paper.

% \end{appendices}

\bibliography{refs}

\end{document}